# Understanding the Relationship Between Personal Data Privacy Literacy and Data Privacy Information Sharing by University Students

Brady D. Lund, Bryan Anderson, Ana Roeschley, Gahangir Hossain

**Abstract**

With constant threats to the safety of personal data in the United States, privacy literacy has become an increasingly important competency among university students, one that ties intimately to the information sharing behavior of these students. This survey-based study examines how university students in the United States perceive personal data privacy and how their privacy literacy influences their understanding and behaviors. Students' responses to a privacy literacy scale were categorized into high and low privacy literacy groups, revealing that high-literacy individuals demonstrate a broader range of privacy practices, including multi-factor authentication, VPN usage, and phishing awareness, whereas low-literacy individuals rely on more basic security measures. Statistical analyses suggest that high-literacy respondents display greater diversity in recommendations and engagement in privacy discussions. These findings suggest the need for enhanced educational initiatives to improve data privacy awareness at the university level to create a better cyber safe population.

# Introduction

Though often a focus of scholarly and public debate, the concept of privacy is rarely defined in the United States context. Though incidents about data privacy infringements are reported on daily, in order to live their day to day lives, Americans "are constantly required to share their data" and though they may be "presented with consent forms full of legalese constantly, opportunities to opt out are rare," further adding to nebulous understandings of privacy and individual behaviors (Roeschley & Frederick, 2024, p. 5). Further complicating these issues in the United States is the fact that privacy is not explicitly enshrined in the Bill of Rights but are rather assumed.

Additionally, privacy laws have not been up for serious debate in Congress for decades, with the Privacy Act of 1974 and the Health Insurance Portability and Accountability Act of 1996 (HIPPA) as the last two major federal privacy laws passed. In the age of big data, major regulations and laws to protect individual privacy are often led by governments outside of United States, as can be seen in the case of the European Union's General Data Protection Regulation (GDPR). Thus, United States citizens often only benefit from privacy protections when corporations find it unprofitable to selectively comply and thus put protections in place for all users internationally. Because major privacy regulations are debated by foreign governments, they are less likely to feature in the news cycles and realms of public debate in the United States. Nonetheless, while privacy regulations have not been enacted on a wide scale in the United States, the exploitation of personal data and information has become a major issue as data breaches and identity scams become a part of everyday life (Cheng et al., 2017; Irvin-Erickson, 2024).

What are the implications of these privacy incongruities for young adults in the United States who are entering adulthood and the responsibilities that come with it? While individuals born after the passage of HIPPA have not been subject to major policy debates regarding privacy in the United States, they still need to have privacy literacy to protect their private data from exploitation. This study uses an open-ended survey question approach to explore how young adults in the United States understand personal data privacy and how their personal data privacy literacy skills affect their abilities to describe responsible privacy behaviors.

# Literature Review

## Understanding Personal Data Privacy Literacy through Statistical Approaches

Privacy literacy is seen as a significant factor when it comes to digital competence. That is because it includes an individual's knowledge, skills, and awareness as it pertains to personal data protection, digital risks, and the implications of data-sharing behaviors (Park, 2013). Existing research has revealed that students who have high privacy literacy skills are more likely to be involved in online protective behaviors such as being able to adjust privacy settings, creating strong passwords, and identifying phishing attempts (Taneja et al., 2014). Research has also shown that privacy literacy levels can differ among various diverse demographic groups. Younger students who have grown up engaged on the internet and other modern information technologies, may sometimes demonstrate overconfidence in their knowledge while lacking an in-depth understanding of security measures (Bartsch & Dienlin, 2016).

Over the years, researchers have developed various methods to measure privacy literacy, including self-assessment surveys, behavioral analysis, and knowledge tests (Masur, 2020). Given the fact that privacy behaviors can be difficult to analyze, using strong statistical approaches is necessary to identify patterns and gaps.

This study uses Shannon's Diversity Index to measure the variation in advice given by students who have different literacy level skills. Previous research (Trepte et al., 2022), has demonstrated how diversity effectively reflects the way individuals with different levels of knowledge express privacy concerns. In addition, previous studies have used methods such as permutation analysis and chi-square tests to assess differences in privacy-related behaviors among certain groups (Taneja et al., 2014). The current study builds upon this work by including these techniques to examine the depth and variety of privacy advice given by university students.

**The Role of Peer Influence**

Privacy literacy skills not only influence how students protect their own data but also how they communicate recommendations for privacy to others. Research suggests that students with higher privacy literacy skills often assume their peers already understand basic privacy principles, leading them to provide less detailed or overly technical advice (Bartsch & Dienlin, 2016). On the other hand, students with lower privacy literacy skills may share suggestions based on their perception of the online audience and social expectations (Litt & Hargittai, 2016). This suggests that privacy decision-making is often guided by social influence rather than technical knowledge, highlighting the need for educational strategies that address both gaining knowledge and behavioral motivations. Furthermore, knowing how privacy literacy not only affects personal behaviors, but also advice-sharing practices are essential for developing more effective privacy education initiatives that can improve both learning and communication strategies.

Research on general user acceptance of advice regarding cybersecurity found that "motivations [for following advice] were predominately individual rather than social" (Fagan & Khan, 2018, p. 32). However, when examining the advice acceptance behaviors among students, found that rather than following advice given in university cybersecurity courses and workshops, such "training is almost marginal for respondents in comparison to internet and friends" for university students in Poland (p. 1279). In the United States context, Conetta (2019) argues that students often "reach out to their social connections in an effort to obtain the most updated and effective [cybersecurity behavior] advice" (p. 16).

**The Privacy Mismatch: Discrepancies Between Knowledge and Behavior**

Privacy literacy plays a fundamental role in shaping an individual's ability to adopt stronger security habits. Milne et al. (2017) found that individuals with higher privacy literacy skills are more likely to be involved in implementing better security practices, such as using multi-factor authentication and avoiding public Wi-Fi networks. In a similar study, Debatin et al. (2009) examined how users managed their online privacy and found a significant gap between awareness and actual practices. In this dilemma, where individuals acknowledge that there are privacy risks but fail to take sufficient security measures, has been extensively documented in the field of cybersecurity (Acquisti et al., 2015).

Furthermore, despite being more knowledgeable when it comes to privacy best practices, students with high privacy literacy skills do not always provide comprehensive privacy advice to others. Bartsch and Dienlin (2016) suggest that while these individuals understand privacy risks, they may fail to effectively communicate due to assumptions that others share their level of knowledge. Nonetheless, Kraus et al. (2023) did find that "students are willing to disseminate [clear privacy literacy training] among fellow students, family members, friends, and colleagues" (p. 241). These findings suggest that privacy education should focus not only on improving personal protective behaviors but also strengthening how individuals share privacy knowledge with others. Kraus et al. found that both student "users and experts struggle to prioritize" cybersecurity advice they receive, a dilemma which is further complicated by the reality that "many representative security awareness websites do not offer a structured way of conveying advice to end users" (p. 237). Without effective knowledge sharing, even those with strong privacy literacy skills may fail to positively influence the privacy practices of their peers.

**Research Gaps in Educational Interventions and Policy Recommendations**

While prior research has explored privacy literacy and protective behaviors, few studies have examined how literacy levels impact the specific privacy advice students share with their peers. There is a limit to understanding how social influence interacts with privacy knowledge and whether those with higher literacy effectively convert their expertise into actionable recommendations. Prior research that explores privacy discrepancies where individuals recognize risks but do not always act accordingly illustrates that only a few have investigated how these inconsistencies apply to advice giving behaviors (Acquisti et al., 2015). This study seeks to address these gaps by analyzing how different levels of privacy literacy impact the content and quality of privacy advice shared among university students. By categorizing students into high and low literacy groups and assessing the diversity and effectiveness of their advice, this research provides insights into how privacy literacy influences knowledge-sharing behaviors.

Knowing that there are inconsistencies as it pertains to privacy knowledge and behavior, scholars are encouraging for ways to improve education for digital privacy literacy. Universities can play an important role in advocating privacy awareness through workshops, digital literacy courses, and the integration of privacy education across various academic disciplines (Youn, 2009). Additionally, Yan et al. (2018) found that most university students tend to have some cybersecurity knowledge, with a need for targeted cybersecurity training for students who have less privacy knowledge. Similarly, Eliza et al. argue that universities "can design and implement more targeted training programs to improve cybersecurity literacy" through identifying the level of cybersecurity awareness among students" (p. 11).

Studies have demonstrated that having structured privacy education programs can lead to improved protective behaviors among students (Solove, 2013). For example, programs that teach students how to recognize phishing attacks, manage privacy settings, and recognize data collection practices have been shown to increase security along with being more conscious when it comes to decision making (Livingston et al., 2019). Prior research also indicates that is important to understand the "correlations between individual differences and security behaviors in order to continue developing the security community's understanding of users" (Gratian et al., 2018, p. 10).

Furthermore, effective policy measures can help bridge the gap between knowledge and action (Lund, 2022). Increased knowledge can be enabled by "uniform data security and data breach notification standards across the US and federal enforcement of these standards" (Irvin-Erickson, 2024, p. 15). Additionally, policymakers have stressed the need for clearer and more transparent privacy policies that enable users to make better informed decisions (Solove, 2013). Although technical measures should be an area of focus, privacy education should also be incorporated to address ethical concerns and the societal impact of data sharing (Nissenbaum, 2010). The current study supports these recommendations by highlighting the gaps in privacy literacy and their implications for information sharing behaviors.

Building upon the existing literature, this study aims to examine how different levels of privacy literacy influence the privacy advice shared among university students. The study is designed to address the following research question:

*How does personal data privacy literacy influence the diversity and depth of privacy-related knowledge, behaviors, and advice shared among university students?*

By employing a mixed-methods approach, the research will analyze both the content and diversity of privacy recommendations given by students with differing literacy levels. The following section outlines the methodological framework used to investigate these relationships, including data collection, participant selection, and analytical techniques.

## Methods

An online survey examining personal data privacy literacy and practices was distributed in late-Fall 2024/early-Spring 2025 to students enrolled in a large, United States-based university. Given the length of the survey, which contained 30 questions, including three open-ended questions, only 90 viable responses were received; however, this number still provides a meaningful sample for analysis of qualitative comments. The primary focus of this research paper is the responses to these three open-ended questions:

- In one to two sentences, or as a list, please describe what advice or lessons on protecting your privacy you have been given since you started college.
- In one to two sentences, or as a list, please describe what advice or lessons on protecting your privacy you have given your friends or family members.
- Please note anything else regarding data privacy that you would like us to know.

The responses to this question were coded into logical categories by human coders, and the frequencies for each category were calculated.

Personal data privacy literacy scores were calculated based on responses to 27 questions related to data privacy. These include behavioral questions such as, "I optimize my privacy settings when I create an online profile," attitudinal questions such as, "Companies seeking information should disclose the way the data are collected, processed, and used," and knowledge questions such as, "Personal information may be used to victimize people." Each of these questions had five responses, which were scored on a scale from 1 to 5 based on the best practices and current research existing in the professional and scholarly literature. Respondents to the survey were sorted into "high" and "low" literacy groups based on their personal data privacy literacy score, which was calculated as a value between 1 and 5. With 90 complete responses to the survey, the 45 respondents with the highest scores were placed in the "high" literacy group, while the other 45 were placed in the "low" literacy group. These groupings were then used to analyze differences in the responses provided to the three open-ended questions, to determine if there is any variance in the personal data privacy advice provided by those with higher or lower literacy in this area.

Several forms of data analysis were performed based on the responses received among the "high" and "low" data privacy literacy groups. The Shannon's Diversity Index was used to assess the variety of advice given by both groups. Permutation analyses and chi-square tests were employed to examine differences in the distribution of responses between the two groups. T-tests were also conducted to analyze the average number of responses provided by each group. These tests were performed so that not just would the actual statements provided by respondents be included, but also analysis of the diversity and frequency of responses between the "high" and "low" literacy groups.

## Results

The following section analyzes the student responses to each of the three open-ended questions based on their personal data privacy literacy score. Sub-headings provide the text of each question that was asked to the respondents.

**"In one to two sentences, or as a list, please describe what advice or lessons on protecting your privacy you have been given since you started college."**

**Table 1. Comparison of Advice or Lessons Gained on Protecting Privacy**

| High Literacy Group | Freq | Low Literacy Group | Freq |
|---|---|---|---|
| Strong, unique passwords | 14 | Strong, unique passwords | 17 |
| Two/Multi factor authentication | 12 | Two/Multi factor authentication | 13 |
| Caution about phishing attacks | 10 | I don't have any guidance | 11 |
| I don't have any guidance | 8 | Avoid sharing any personal info online | 8 |
| Avoid sharing any personal info online | 7 | Caution about phishing attacks | 6 |
| Avoid sharing too much info on social media | 6 | Use a VPN | 5 |
| Do not share passwords | 4 | Review privacy settings regularly | 4 |
| Use a VPN | 4 | Do not share passwords | 4 |
| Review privacy settings regularly | 4 | Avoid public Wi-Fi when possible | 4 |
| Avoid public Wi-Fi when possible | 3 | Avoid sharing too much info on social media | 2 |
| Regularly update software | 3 | Regularly update passwords | 1 |
| Regularly update passwords | 2 | | |
| Use Google authenticator | 1 | | |
| Verify the authenticity of links | 1 | | |
| Check what cookies are being collected online | 1 | | |
| Avoid sharing personal info in any format | 1 | | |
| Avoid apps with known data leaks | 1 | | |

The guidance provided among the respondents to this survey with high personal data privacy literacy differs in both depth and variety compared to that of students with lower literacy levels. Students in the high literacy group gave more diverse recommendations (Shannon's Diversity Index = 2.51) compared to the low literacy group (Index = 2.21), suggesting that these respondents may have a broader understanding of privacy protection measures. Their responses mentioned a variety of advanced security practices such as using Google Authenticator, verifying the authenticity of links, checking cookies collected online, and avoiding apps with known data leaks. These "high literacy" students also emphasized using strong, unique passwords (14 mentions) and enabling two/multi-factor authentication (12 mentions), alongside general advice like avoiding public Wi-Fi and regularly updating software.

Students with lower data privacy literacy, conversely, appear to focus more on fundamental security practices. Their top suggestions included using strong, unique passwords (17 mentions), implementing two/multi-factor authentication (13 mentions), and avoiding sharing personal information online (8 mentions). However, this group had a significant proportion (11 mentions) of students who reported not having any guidance. The lower literacy group also provided fewer advanced recommendations, with little mention of more nuanced topics like cookie management or link verification. The permutation analysis conducted on this data (p = .099) shows some evidence of greater diversity in advice among the high literacy group, although the chi-square test (p = 0.794) indicates no statistically significant difference in the overall distribution of responses.

**"In one to two sentences, or as a list, please describe what advice or lessons on protecting your privacy you have given your friends or family members."**

**Table 2. Comparison of Advice or Lessons Given on Protecting Privacy**

| High Literacy Group | Freq | Low Literacy Group | Freq |
|---|---|---|---|
| Be cautious of suspicious links | 12 | Strong, unique passwords | 9 |
| Avoid sharing personal info on social media | 11 | I don't have any guidance | 9 |
| Strong, unique passwords | 10 | Be cautious of suspicious links | 8 |
| Be cautious of suspicious emails | 10 | Be cautious of suspicious emails | 7 |
| Use two-factor authentication | 9 | Use two-factor authentication | 5 |
| Do not share personal info with anyone | 7 | Avoid sharing personal info on social media | 3 |
| Avoid using public Wi-Fi if possible | 5 | Avoid using public Wi-Fi if possible | 3 |
| I don't have any guidance | 4 | Check for spelling errors for phishing | 2 |
| Do not share passwords | 4 | Always read the fine print of privacy policies | 2 |
| Regularly update passwords | 3 | Do not share passwords | 2 |
| Utilize VPNs | 2 | Nothing online is actually "free" | 1 |
| Always read the fine print of privacy policies | 2 | Use a password manager | 1 |
| Do not save passwords on your device | 1 | Use a fake name online | 1 |
| Use a password manager | 1 | Do not use websites if unnecessary | 1 |
| Rely only on trusted sources of info | 1 | Turn off location capturing on photos | 1 |
| Always be skeptical | 1 | Disable unnecessary cookies | 1 |
| Disable unnecessary cookies | 1 | | |
| Do not transfer money online except to trusted people | 1 | | |
| Cross-verify the authenticity of websites | 1 | | |
| Don't worry, be happy | 1 | | |

Evident in the analysis of advice given by students with high and low personal data privacy literacy to friends and family are some notable differences in both the depth and diversity of recommendations. Students in the high literacy group provided a broader range of advice (Shannon's Diversity Index = 2.77) compared to the low literacy group (Index = 2.50), indicating that these students may possess a more comprehensive understanding of privacy protection practices. This group often offered advice on advanced privacy measures such as cross-verifying website authenticity, disabling unnecessary cookies, and using VPNs. The most common recommendations provided include avoiding sharing personal information on social media (11 mentions), being cautious of suspicious links (12 mentions), and using strong, unique passwords (10 mentions). These respondents also demonstrated a proactive approach to data privacy, with suggestions like regularly updating passwords and not saving passwords on devices.

Students with lower privacy literacy comparatively focused more on fundamental and easily understandable practices. Their top suggestions included using strong, unique passwords (9 mentions), being cautious of suspicious emails (7 mentions), and enabling two-factor authentication (5 mentions). However, this group also had a significant portion (9 mentions) who reported not providing any privacy guidance, suggesting a lower level of engagement with the topic. While some advanced suggestions were made, such as disabling unnecessary cookies and using fake names online, these were far less common. The permutation analysis ($p = .208$) and chi-square test ($p = .198$) indicate no statistically significant difference in the overall distribution of responses. However, the t-test ($t = 6.46$, $p < .01$) shows a significant difference in the number of responses provided, with the high literacy group offering substantially more advice on average (1.93 responses per person) than the low literacy group (1.24 responses per person). This may suggest that higher literacy students not only possess greater knowledge about data privacy but are also more likely to share this knowledge with others.

**"Please note anything else regarding data privacy that you would like us to know."**

**Table 3. Comparison of Additional Comments Provided**

| High | Freq | Low | Freq |
|---|---|---|---|
| No comments | 30 | No comments | 36 |
| Changes to data privacy are needed | 3 | We must remain vigilant in preserving our privacy | 2 |
| Privacy is a myth | 2 | The current generation is becoming more apathetic to security risks | 1 |
| We must develop better security measures | 1 | Two step verification | 1 |
| Education about password managers must be expanded | 1 | How do I know if companies are using my data? | 1 |
| We need better education about data privacy best practices | 1 | We can make informed decisions about whether to use the Internet, knowing our data is at risk of being sold | 1 |
| Privacy policies need to be more clear and concise | 1 | Changes to data privacy are needed | 1 |
| Old accounts that are no longer in use should be deleted | 1 | It is important to have these conversations with family and friends | 1 |
| Personal privacy impacts collective security | 1 | Encryption is needed | 1 |
| We need to educate about the digital footprint | 1 | | |
| Privacy of personal data is more important than accuracy | 1 | | |
| Don't track my data! | 1 | | |
| I have been a victim of phishing | 1 | | |

The analysis of responses to the question about additional thoughts on data privacy shows some very clear differences between students with high and low data privacy literacy. The high literacy group demonstrated significantly greater diversity in their responses (Shannon's Diversity Index = 1.58) compared to the low literacy group (Index = 1.17), with the permutation analysis (p = .035) confirming this difference. The high literacy group expressed a wide range of concerns and suggestions, including the need for better education on data privacy best practices, clearer privacy policies, and expanded knowledge about password managers. Other noteworthy comments included the importance of deleting unused accounts, the collective impact of personal privacy on security, and skepticism reflected in the statement "Privacy is a myth." Although 30 respondents from this group had no additional comments, the breadth of topics discussed by the remaining respondents highlights a deeper engagement with the complexities of data privacy.

Responses from the low literacy group were more concentrated, with 36 respondents offering no additional comments. The few responses that were provided focused on general concerns and basic concepts. These included the need for two-step verification, maintaining vigilance in protecting privacy, and questioning how companies use personal data. Some responses touched on broader societal themes, such as concerns about generational apathy toward security risks and the importance of discussing privacy with family and friends. However, the overall range of topics discussed by the low literacy group was narrower. The chi-square test ($X^2 = 44.18$, $p < .01$) further supports a statistically significant difference in the distribution of responses between the two groups.

## Discussion

The findings from this study support prior research that suggest that privacy literacy and information sharing behaviors can affect the role of social influence and communication clarity when it comes to guiding privacy decisions. Students that have high privacy literacy skills provide more technically advanced recommendations, although the advice was less detailed, suggesting they assumed that their peers had somewhat near the same level of understanding (Bartsch & Dienlin, 2016). Meanwhile, students with low literacy skills were more willing to share basic, easy to understand privacy recommendations, which were a result of social influences and anticipated responses from their peers (Litt & Hargittai, 2016). This aligns with prior studies on privacy decision-making, that highlights that students may adjust their privacy habits based on social influence rather than specific technical knowledge (Trepte et al., 2022).

While the presence of these sentiments is higher in the high literacy respondents group, responses from both groups indicate the presence of cynicism and distrust regarding privacy (e.g., "privacy is a myth"). This cynicism is explored by Roeschley and Frederick (2024) who argue that despite their ubiquity, "in the age of networked information and big data, consent forms are designed to devalue consent" and erase notions of privacy as a human right (p. 5). These sentiments are echoed by Nicholson et al. (2021) who found that young people participating in a cybersecurity workshop "reported an expectation that a security compromise would be imminent in their future, and as a consequence they were more relaxed about using security tools or enacting security behaviors – similar to arguments made for privacy fatigue" (p. 208).

That the presence of cynicism in this study is expressed by members of both literacy groups indicates broad public knowledge of personal privacy violations among students. Yet, an awareness of privacy breaches does not translate to behaviors that result in personal data privacy protections. Irvin-Erickson (2024) argues that "there is a need for individuals, law enforcement, victim service providers, and policymakers to put as much emphasis on the acquisition of personal information as the subsequent frauds" (p. 14). These findings propose that education on privacy literacy should focus on strengthening student's ability to communicate privacy knowledge and improving technical literacy skills more effectively. Without clear and user-friendly guidance for privacy recommendations, even well-informed individuals could fail to better influence the privacy habits of their peers.

It is also important to address the ways in which expressions of cynicism suggest that individuals feel like their privacy behaviors have little impact on their (lack of) privacy. This should be of concern to data privacy researchers and educators, as it suggests a breach of a societal trust on a macro level. On a more micro and institutional level, cynicism among students should raise concerns because it can prevent individual students in engaging in behaviors that to secure both their own individual data and consequently, their institutional data. Jones et al. (2023) argue that institutions of higher ed themselves actively contribute to the disempowerment of their students when they "aggregate, analyze, and use student data for seemingly innumerable purposes" (p. 486). In order to combat this exploitation, Roeschley and Frederick argue that "rather than participate in the constant surveillance of our students" higher ed faculty should foster critical thought "about consent and data exploitation by refusing to utilize proctoring software in class assignments, by explicitly asking for student consent when we use their data, and by being transparent about what drives our decisions" (p. 5). Though these steps may be seen as contrary to institutional interests, they may be necessary to enabling student empowerment and motivating responsible privacy behaviors. After all, if as Nicholson et al. suggest, young people feel so disempowered that they believe privacy breaches to be inevitable, why would they take the time to engage in safe behaviors? Why would they protect their own privacy, much less the university's?

## Conclusion

This study found both differences between students with high personal data literacy and students with low literacy. High literacy respondents demonstrated a proactive approach to data privacy, with suggestions like regularly updating passwords and not saving passwords on devices. Meanwhile, students with lower privacy literacy comparatively focused more on easily understandable practices and indicated an overall lower level of engagement with the topic of personal data privacy. Interestingly, members of both groups expressed a sense of cynicism regarding privacy. This indicates that in order to ensure that the presence of privacy literacy results in privacy behaviors, a serious investment from the educational and policy sectors in breaking through cynicism and privacy fatigue is necessary.

Nevertheless, the results of this study indicate that higher personal data literacy students not only possess greater knowledge about data privacy but are also more likely to share this knowledge with others. Future research should explore how privacy education programs could implement peer-led learning models to improve both student privacy literacy and the ability to convey privacy knowledge in significant ways. Furthermore, the statistical difference in privacy-related responses between high and low literacy groups suggests the need for tailored strategies that focus on bridging the communication and research gaps regarding personal data privacy.

# References

Acquisti, A., Brandimarte, L., & Loewenstein, G. (2015). Privacy and human behavior in the age of information. *Science, 347*(6221), 509-514. https://doi.org/10.1126/science.aaa1465

Bartsch, M., & Dienlin, T. (2015). Control your Facebook: An analysis of online privacy literacy. *Computers in Human Behavior, 56*(March 2016), 147-154. https://doi.org/10.1016/j.chb.2015.11.022

Cheng, L., Liu, F., & Yao, D. (2017). Enterprise data breach: Causes, challenges, prevention, and future directions. *Wiley Interdisciplinary Reviews: Data Mining and Knowledge Discovery*, *7*(5), e1211. https://doi.org/10.1002/widm.1211

Conetta, C. (2019). Individual differences in cyber security. *McNair Research Journal SJSU*, *15*(1), 4. https://doi.org/10.31979/mrj.2019.1504

Debatin, B., Lovejoy, J. P., Horn, A. K., & Hughes, B. N. (2009). Facebook and Online Privacy: Attitudes, behaviors, and unintended consequences, *Journal of Computer-Mediated Communication*, *15*(1), 83–108. https://doi.org/10.1111/j.1083-6101.2009.01494.x

Eliza, F., Fadli, R., Ramadhan, M. A., Sutrisno, V. L. P., Hidayah, Y., Hakiki, M., & Dermawan, D. D. (2024). Assessing student readiness for mobile learning from a cybersecurity perspective. *Online Journal of Communication and Media Technologies*, *14*(4), e202452. https://doi.org/10.30935/ojcmt/15017

Fagan, M., & Khan, M. M. H. (2018). To follow or not to follow: A study of user motivations around cybersecurity advice. *IEEE Internet Computing*, *22*(5), 25-34.

Gratian, M., Bandi, S., Cukier, M., Dykstra, J., & Ginther, A. (2018). Correlating human traits and cyber security behavior intentions. *computers & security*, *73*, 345-358.

Irvin-Erickson, Y. (2024). Identity fraud victimization: A critical review of the literature of the past two decades. *Crime Science*, *13*(1), 3. https://doi.org/10.1186/s40163-024-00202-0

Jones, K. M., Goben, A., Perry, M. R., Regalado, M., Salo, D., Asher, A. D., ... & Briney, K. A. (2023). Transparency and Consent: Student Perspectives on Educational Data Analytics Scenarios.portal: Libraries and the Academy,23(3), 485-515.

Kraus, L., Švábenský, V., Horák, M., Matyás, V., Vykopal, J., & Celeda, P. (2023, June). Want to Raise Cybersecurity Awareness? Start with Future IT Professionals. In *Proceedings of the 2023 Conference on Innovation and Technology in Computer Science Education V. 1* (pp. 236-242). https://doi.org/10.1145/3587102.3588862

Litt, E., & Hargittai, E. (2016). The imagined audience on social network sites. *Social Media + Society, 2*(1). https://doi.org/10.1177/2056305116633482

Livingstone, S., Stoilova, Mariya., & Nandagiri, R. (2019). *Children's data and privacy online: Growing up in a digital Age: An evidence review.* https://www.lse.ac.uk/media-and-

communications/assets/documents/research/projects/childrens-privacy-online/Evidence-review.pdf

Lund, B. (2022). Policy Before Technology: Don't Outkick the Coverage. *Information Technology and Libraries, 41*(1). https://doi.org/10.6017/ital.v41i1.14773

Masur, P. K. (2019). *Situational privacy and self-disclosure: Communication processes in online environments.* Springer International. https://doi.org/10.1007/978-3-319-78884-5

Milne, G. R., LaBrecque, L. I., & Cromer, C. (2009). Toward an understanding of the online consumer's risky behavior and protection practices. *Journal of Consumer Affairs, 43*(3), 449-473. https://doi.org/10.1111/j.1745-6606.2009.01148.x

Nicholson, J., Terry, J., Beckett, H., & Kumar, P. (2021, October). Understanding young people's experiences of cybersecurity. In *Proceedings of the 2021 European Symposium on Usable Security* (pp. 200-210).

Nissenbaum, H. (2009). *Privacy in context: Technology, policy, and the integrity of social life.* Stanford University Press. https://doi.org/10.1515/9780804772891

Park, Y. J. (2011). Digital literacy and privacy behavior online. *Communication Research, 40*(2), 215-236. https://doi.org/10.1177/0093650211418338

Roeschley, A., & Frederick, M. D. (2024). The Difference Between Passive Subjects and Autonomous Actors: Proposing an Orientation Toward Consent in Library and Information Science. In *Proceedings of the ALISE Annual Conference*. https://doi.org/10.21900/j.alise.2024.1726

Solove, D. J. (2013). Privacy self-management and the consent dilemma. *Harvard Law Review 1880, 126*(7). https://harvardlawreview.org/print/vol-126/introduction-privacy-self-management-and-the-consent-dilemma/

Szumski, O. (2018). Cybersecurity best practices among Polish students. *Procedia Computer Science*, *126*, 1271-1280. https://doi.org/10.1016/j.procs.2018.08.070

Taneja, A., Vitrano, J., & Gengo, N. J. (2014). Rationality-based beliefs affecting individual's attitude and intention to use privacy controls on Facebook: An empirical investigation. *Computers in Human Behavior, 38*(September 2024), 159-173. https://doi.org/10.1016/j.chb.2014.05.027

Trepte, S. (2021). The social media privacy model: Privacy and communication in the light of social media affordances. *Communication Theory, 31*(4), 549-570. https://doi.org/10.1093/ct/qtz035

Yan, Z., Robertson, T., Yan, R., Park, S. Y., Bordoff, S., Chen, Q., & Sprissler, E. (2018). Finding the weakest links in the weakest link: How well do undergraduate students make cybersecurity judgment?. *Computers in Human Behavior*, *84*, 375-382. https://doi.org/10.1016/j.chb.2018.02.019

Youn, S. (2009). Determinants of online privacy concern and its influence on privacy protection behaviors among young adolescents. *Journal of Consumer Affairs, 43*(3), 389-418. https://doi.org/10.1111/j.1745-6606.2009.01146.x